\documentstyle[aps]{revtex}
\def\gfi{\stackrel{5}{g}}
\def\gfo{\stackrel{4}{g}}
\def\Rfi{\stackrel{5}{R}}

\begin{document}
%\twocolumn[\hsize\textwidth\columnwidth\hsize\csname @twocolumnfalse\endcsname
\title{Brane-world cosmology}
\author{Daisuke Ida}
\address{
Department of Physics, Kyoto University, Kitashirakawa,
Sakyo-ku, Kyoto 606-8502, Japan\\ 
{\rm E-mail address: ida@tap.scphys.kyoto-u.ac.jp}\\
}
\maketitle
\begin{abstract}
A simple model of the brane-world cosmology has been proposed,
which is characterized by four parameters, the bulk
 cosmological constant, 
the spatial curvature of the universe,
the radiation strength arising from bulk space-time
and the breaking parameter of $Z_2$-symmetry.
The bulk space-time is assumed to be locally static 
five-dimensional analogue of 
the Schwarzschild-anti-de Sitter space-time, and then
the location of three-brane is determined by metric 
junction. The resulting Friedmann equation recovers
standard cosmology, and a new term arises if the assumption
$Z_2$-symmetry is dropped, 
which behaves as cosmological term
in the early universe, next turns to negative curvature
term, and finally damps rapidly.
\end{abstract}
%]
\medskip
Recently, Randall and Sundrum have proposed two
 models in which
our universe is a three-brane imbedded in a five-dimensional
anti-de Sitter space $AdS_5$ as a possible solution to the
hierarchy problem between weak and Planck scales. 
In their first model, the fifth-dimension is bounded
by two domain walls with positive and negative tension,
and the large mass hierarchy can be explained
in terms of the exponential factor of the $AdS_5$ 
metric along the extra dimension. In this model, 
the visible world is assumed to be negative-tension brane,
however, Shiromizu et al. have shown that the gravitation
is repulsive on such a brane, so that physically
inacceptable. 
In second model, they suggested that the extra dimension
need not be compact, and that
 it is sufficient to put a single positive-tension brane.
They have also shown that zero-mode of the linear metric
perturbation recovers Newton's inverse-square law and that
the Kaluza-Klein modes give only small contribution to the
gravitaion on the brane.
Currently, there seems to be no strong evidence excluding
this model.
On the other hand, it is also of theoretical 
interest to seek for the cosmological and astrophysical
prediction of this brane-world senario. 
Especially, the black holes \cite{CHR,EHM}
and expanding universe \cite{BDL}-\cite{M}
have been researched by many authors.
As far as a black hole solution, there is only known for
(2+1)+1-dimensional bulk space-time \cite{EHM}, so that
a (3+1)+1-dimensional solution is desired.
While for the expanding universe, many authors have 
derived the evolution equation for the metric variables,
there are a few versions of explicit bulk solution.
Moreover, the known bulk solution has a rather complecated
expression, it seems to be difficult to handle.
To test the brane-world scenario observationally, it would 
be necessary to know the 
evolution of density fluctuation.
The density fluctuation cannot be determined only 
by the background metric of three-brane, but we should know the 
five-dimensional metric also, since the electric part of five-dimensional
Weyl tensor contributes to the gravitational field on the three-brane
through the four-dimensional Einstein equation \cite{SMS}. 
However, as mensioned above, though Friedmann equations have been derived for several cases,
only a few global solutions are known, so that we shall
attempt to construct a simple model of brane-world 
cosmology.
The usual approach has been as follows:
 assume the appropriate
form of the metric which is {\em manifestly $Z_2$-symmetric},
and solve the junction condition and the Einstein equation.
Alternatively, here we assume an appropriate five-dimensional metric at
the beginning, and then find the location of three-brane.

Let us consider the five-dimensional bulk space-time.
In the original Randall-Sundrum scenario, they take 
$AdS_5$ as the bulk space-time,
here we assume that is to be Einstein space with negative cosmological
constant. As such space, we take the five-dimensional analogue of the
Schwarzschild-anti-de Sitter space-time in the following form
\begin{equation}
\gfi =-h(a)dt^2+\frac{1}{h(a)}da^2+a^2\left[
d\chi^2+f_k(\chi)^2(d\vartheta^2+\sin^2\vartheta d\varphi^2)\right],
~~~(k=0,\pm1)
\end{equation}
where $f_0(\chi)=\chi$, $f_1(\chi)=\sin\chi$ and
$f_{-1}(\chi)=\sinh\chi$.
The solution of field equation
$\Rfi_{\mu\nu}=\stackrel{5}{\Lambda}\gfi_{\mu\nu}$
with the cosmological constant
$\stackrel{5}{\Lambda}=-4l^2$ is given by
\begin{equation}
h(a)=k-\frac{\alpha}{a^2}+l^2 a^2
\end{equation}
where $\alpha$ and $l$ are constants. If $\alpha=0$, then this reduces to
$AdS_5$, however, $\alpha\ne0$ generates the electric part of Weyl
tensor.
The three-brane model can be obtained in the similar manner to the spherical
thin-shell model in four-dimension \cite{I}, i.e., by patching together two
manifolds, which may have a different parameter $\alpha$ from each other,
but with common $k$ and $l$, at the timelike hypersurface.

We shall consider the location of three-brane in the form
$t=t(\tau)$, $a=a(\tau)$ parametrized by the proper time
$\tau$ on the brane, then the induced metric of three-brane
will be given by
\begin{equation}
\gfo=-d\tau^2+a(\tau)^2\left[d\chi^2+f(\chi)^2
(d\vartheta^2+\sin^2\vartheta\varphi^2)\right],
\end{equation}
namely, $\tau$ and $a(\tau)$ correspond to the cosmic time
and scale factor of Friedmann-Robertson-Walker universe,
respectively.
%For concreteness, we assume that the brane is composed of
%the vacuum energy, and that the bulk space is refrection
%symmetric with respect to the brane.
The tangent vector of the brane can be written in the form
\begin{equation}
u=\dot {t}_{R,L}\frac{\partial}{\partial {t_{R,L}}}
+\dot a\frac{\partial}{\partial a},
\end{equation}
where the dot denotes the derivative with respect to a proper time $\tau$,
and the subscripts $R$ and $L$ denote the quantity evaluated 
on the right and the left side of the brane, respectively
(note that the scale factor $a(\tau)$ and the proper time $\tau$
are common on both sides).
Given the tangent vector $u$, the normal 1-form 
to the three-brane $n$ is determined
up to the sign. Here we fix this freedom by setting
\begin{eqnarray}
n&=&\dot a dt_R-\dot t_R da,\nonumber\\
&=&-\dot a dt_L+\dot t_L da.
\end{eqnarray}
This choice gives the Randall-Sundrum model in the limit $\alpha_{R,L}
\rightarrow 0$ (other choices are possible, though the resulting
Friedmann equation is not affected).

The embedding of three-brane is characterized by the induced metric $\gfo$ and
the extrinsic curvature $K_{R,L}$ on both sides
\begin{eqnarray}
&&\gfo_{\mu\nu}=\gfi_{\mu\nu}-n_\mu n_{\nu},\\
&&K_{\mu\nu}=\gfo_{\mu}{}^{\lambda}\nabla_{\lambda}n_{\nu},
\end{eqnarray}
where $\nabla$ denotes the Riemannian connection compatible with
$\gfi$. The extrinsic curvature has the following non-vanishing components
\begin{eqnarray}
&&(K_R)_{\mu\nu}u^{\mu}u^{\nu}=
(h_R\dot t_R)^{-1}(\ddot a+h_R'/2),\\
&&(K_R)_{\chi}^{\chi}=(K_R)_{\vartheta}^{\vartheta}=
(K_R)_{\varphi}^{\varphi}=-h_R\dot t_R/a,\\
&&(K_L)_{\mu\nu}u^{\mu}u^{\nu}=-
(h_L\dot t_L)^{-1}(\ddot a+h_L'/2),\\
&&(K_L)_{\chi}^{\chi}=(K_L)_{\vartheta}^{\vartheta}=
(K_L)_{\varphi}^{\varphi}=h_L\dot t_L/a,
\end{eqnarray}
where $h_{R,L}=k-\alpha_{R,L}/a^2+l^2 a^2$, thus the extrinsic curvature
has the jump at the brane. This jump can be interpreted as to be 
generated by the matter field confined to the brane.
The surface stress-energy tensor on the brane $S$ is defined by
\begin{equation}
S_{\mu\nu}=-\frac{1}{\kappa^2}\left\{K_{\mu\nu}-K_{\lambda}
^{\lambda}\gfo_{\mu\nu}\right\}^-,
\end{equation}
where for any tensor field $Q$, $\{Q\}^-=Q_R-Q_L$ is defined.
In the present case, $S$ necessary has the ideal-fluid form
\begin{equation}
S_{\mu\nu}=(\varrho+p)u_{\mu}u_{\nu}+p\gfo_{\mu\nu},
\end{equation}
due to the spatial symmetry of the three-section generated by 
$Span\{d\chi,d\vartheta,d\varphi\}$, where $\varrho$ and $p$
are the energy density and pressure of the brane, 
respectively, or equivalently,
\begin{equation}
\left\{K_{\mu\nu}\right\}^-=-\kappa^2\left(S_{\mu\nu}
-\frac{1}{3}S^{\lambda}_{\lambda}\gfo_{\mu\nu}\right)
=-\kappa^2\left[(\varrho+p)u_{\mu}u_{\nu}+\frac{\varrho}{3}\gfo_{\mu\nu}
\right].\label{def}
\end{equation}
The Gauss-Codazzi equations lead to the following metric junction
conditions \cite{I}
\begin{eqnarray}
&&\frac{\kappa^2}{2}[(K_R)_{\mu\nu}+(K_L)_{\mu\nu}]S^{\mu\nu}
=\left\{\stackrel{5}{G}_{\mu\nu}n^{\mu}n^{\nu}\right\}^-\label{gauss}\\
&&\kappa^2\gfo_{\mu}{}^{\lambda}\nabla_{\nu}S_{\lambda}{}^{\nu}
=\gfo_{\mu}{}^{\lambda}\left\{\stackrel{5}{G}_{\lambda}{}^{\nu}n_{\nu}
\right\}^-\label{conservation}
\end{eqnarray}
%%%%%%%%%%%%%%%%%%%%%%%%%%%%%%%%%
The last equation is the conservation law, which in the present case
leads to a familiar equation
\begin{equation}
\frac{d}{d\tau}(\varrho a^3)+p\frac{d}{d\tau}(a^3)=0.
\end{equation}
While, the explicit forms of
Eqs.~(\ref{def}), (\ref{gauss}) are
\begin{eqnarray}
&&(h_R\dot t_R)^{-1}\left(\ddot a+\frac{h_R'}{2}\right)+
(h_L\dot t_L)^{-1}\left(\ddot a+\frac{h_L'}{2}\right)
%=-\kappa^2\left(p+\frac{n-2}{n-1}\varrho\right),
=-\kappa^2\left(p+\frac{2}{3}\varrho\right),\\
&&h_R\dot t_R+h_L\dot t_L
%=-\kappa^2\frac{\varrho}{(n-1)}a
=\frac{\kappa^2}{3}\varrho a,\\
&&(h_R\dot t_R)^{-1}\left(\ddot a+\frac{h_R'}{2}\right)-
(h_L\dot t_L)^{-1}\left(\ddot a+\frac{h_L'}{2}\right)
%=(n-1)\frac{p}{\varrho a}(h_R\dot t_R+h_L\dot t_L)
=3\frac{p}{\varrho a}(h_R\dot t_R-h_L\dot t_L),
\end{eqnarray}
which lead to
\begin{eqnarray}
&&h_R\dot t_R
%=\frac{(n-1)}{2(n-2)\kappa^2\varrho}(h_R'-h_L')
%-\frac{\kappa^2}{2(n-1)}\varrho a
=-\frac{3}{4\kappa^2\varrho}(h_R'-h_L')
+\frac{\kappa^2}{6}\varrho a\label{tR}\\
&&h_L\dot t_L
%=\frac{(n-1)}{2(n-2)\kappa^2\varrho}(h_R'-h_L')
%+\frac{\kappa^2}{2(n-1)}\varrho a
=\frac{3}{4\kappa^2\varrho}(h_R'-h_L')
+\frac{\kappa^2}{6}\varrho a,
\end{eqnarray}
and the equation of motion
\begin{equation}
\ddot a=
%\frac{(n-1)^3}{4(n-2)^2\kappa^4}\frac{p}{\varrho^3 a}
%(h_R'-h_L')^2-\frac{h_R'+h_L'}{4}
%-\frac{(n-2)}{4(n-1)^2}\kappa^4\varrho
%\left(\varrho+\frac{n-1}{n-2}p\right)a\nonumber\\
%&&=\frac{(n-1)^3(\alpha_R-\alpha_L)^2}{(n-2)^2\kappa^4}
%\frac{p}{\varrho^3 a^7}-\frac{\alpha_R+\alpha_L}{2a^3}
%-\frac{(n-2)\kappa^4}{4(n-1)^2}\varrho
%\left(\varrho+\frac{n-1}{n-2}p\right)a-l^2a
%\frac{27}{16\kappa^4}\frac{p}{\varrho^3 a}
%(h_R'-h_L')^2-\frac{h_R'+h_L'}{4}
%-\frac{\kappa^4}{18}\varrho
%\left(\varrho+\frac{3}{2}p\right)a\nonumber\\=
\frac{27(\alpha_R-\alpha_L)^2}{4\kappa^4}
\frac{p}{\varrho^3 a^7}-\frac{\alpha_R+\alpha_L}{2a^3}
-\frac{\kappa^4}{18}\varrho
\left(\varrho+\frac{3}{2}p\right)a-l^2a. 
\end{equation}
The first integral of this equation, namely the Friedmann
equation, can be obtained by putting Eq. (\ref{tR})
into the normalization condition
$-h_R\dot{t}_R^2+\dot{a}^2/h_R=-1$, and we have 
\begin{equation}
\left(\frac{\dot a}{a}\right)^2
=-\frac{k}{a^2}-l^2+\frac{\kappa^4}{36}\varrho^2
+\frac{\alpha_R+\alpha_L}{2a^4}
+\frac{9(\alpha_R-\alpha_L)^2}{4\kappa^4\varrho^2a^8}.
\label{Friedmann}
\end{equation}
This result is in agreement with Bin\'etruy et al. 
\cite{BDEL} except
the last term presented here. 
This new term arises due to the asymmetry
between the left and right world, which vanishes in the
$Z_2$-symmetric case $\alpha_R=\alpha_L$.
According to the brane-world scenario, it is natural
to assume that the matter component consists of vacuum
energy and the ordinary matter, namely
\begin{eqnarray}
&&\varrho=\rho+\sigma,\\
&&p=P-\sigma,
\end{eqnarray}
where $\rho$ and $P$ denote the energy density and pressure
of ordinary matter, respectively, and
$\sigma={\rm constant}$ being the tension of the brane.
Then the conservation equation (\ref{conservation})
and the Friedmann equation (\ref{Friedmann})
becomes
\begin{eqnarray}
&&\frac{d}{d\tau}(\rho a^3)+P\frac{d}{d\tau}(a^3)=0,\\
&&\left(\frac{\dot a}{a}\right)^2
=\frac{8\pi G_N}{3}\rho
-\frac{k}{a^2}+\frac{\Lambda}{3}
+\frac{\alpha_R+\alpha_L}{2a^4}
+\frac{\kappa^4}{36}\rho^2
+\frac{9(\alpha_R-\alpha_L)^2}{4\kappa^4
(\rho+\sigma)^2a^8},\label{Friedmann2}
\end{eqnarray}
where
\begin{eqnarray}
&&8\pi G_N =\frac{\kappa^4\sigma}{6},\\
&&\Lambda=\frac{\kappa^4\sigma^2}{12}-3l^2,
\end{eqnarray}
are regarded as four-dimensional Newton's constant of
gravitation and the cosmological constant, respectively.
The equation (\ref{Friedmann2})
 recovers standard cosmology in a good 
approximation for appropriate ranges of parameters.
The first three terms on r.h.s. are usual in standard
cosmology. 
The forth term behaves like radiation, however,
this can be negative in the present model.
The origin of this term is now clarified,
i.e., the electric (Coulomb) part of Weyl tensor
in the five-dimensional back ground metric.
The fifth term rapidly damps in usual, e.g. 
$\propto a^{-6}$ for dust,
and $\propto a^{-8}$ for radiation, so that
this term becomes significant in the early universe.
Bin\'etury et al. \cite{BDEL} 
have been discussed a constraint
on this term by nucleosynthesis.
The last term, which presents if the assumption
$Z_2$-symmetric is dropped, behaves in various ways.
Assuming that $\rho\ll \sigma$, it behaves like
positive $\Lambda$-term for radiation, so that 
we might have a inflation without any other fields in the 
early universe. For dust, it behaves like 
negative curvature term, 
and eventually, when $\sigma>\rho$, it will rapidly damp.

Thus, our model seems to give us a many features 
of brane-world cosmology, in fact, there are 6 cosmological
parameters corresponding to 6 terms in r.h.s. of 
Eq.(\ref{Friedmann2}), so that one might applicate this model
to the dark matter, or cosmological constant problem. 
Since the bulk space-time is given in the simplest
form imaginable, it is easy to handle, so that
the calculation of density fluctuation 
or gravitational radiation would be possible, which 
are of interest from the viewpoints of observational
cosmology and astrophysics.

{\em Acknowledgements.}\\
We would like to acknowledge many helpful discussions with 
Prof. Humitaka Sato,  Dr. Tetsuya Shiromizu,
Prof. Ken-ichi Nakao and Dr. Akihiro Ishibashi.

\end{document}